\documentstyle[sprocl]{article}
\def\Journal#1#2#3#4{{#1} {\bf #2}, #3 (#4)}

\def\PRL{\em Phys. Rev. Lett.}

\def\be{\begin{equation}}
\def\ee{\end{equation}}
\def\bea{\begin{eqnarray}}
\def\eea{\end{eqnarray}}

\begin{document}
\title{WHAT IS FOUND UPON DEFROSTING THE UNIVERSE AFTER INFLATION~\footnote{Talk given at the 18th {\it Texas Symposium on Relativisitc Astrophysics}, December 15-20, Chicago, Illinois.}}
\author{ A. RIOTTO }
\address{NASA/Fermilab Astrophysics Center, \\ Fermilab
National Accelerator Laboratory, Batavia, Illinois~~60510-0500, USA}
\maketitle\abstracts{At the end of inflation the universe is frozen in a near zero-entropy
state with energy density in a coherent scalar field and must be
"defrosted" to produce the observed entropy and baryon number.  Baryon asymmetry may be generated by the decay of
supermassive Grand Unified Theory (GUT) bosons produced non-thermally
in a preheating phase after inflation, thus solving many
drawbacks facing GUT baryogenesis in the old reheating scenario.}
\section{Prologo}
Before going into details, let me  explain the origin of the term "defrost" and its derivatives~\cite{paper}. After inflation the universe appears slightly boring: no particles around, zero-entropy density, no thermal bath and all the energy density stored in the inflaton scalar field. It is clear that  the universe
must undergo a sort of phase transition from this state to produce the  observed entropy and a thermal bath of particles. This process was denominated "defrosting" by R. Kolb. However, our collaborator A. Linde felt rather uncomfortable with this denomination. He wrote (quote):"...
It is a cool idea and I am sure that Rocky will use this word in one of his brilliant talks. However, I really have severe problems with it. First of all, I associate it with the frozen chicken breasts, which I cannot stand...". In spite of Andrei's opinion, I decided to adopt the term "defrosting" anyway when Rocky dropped in my office and, with a very strong Chicago-Mafia accent, asked
me how would I have translated the expression "{\it I break your legs}" in italian~\footnote{That sounds something like "{\it Ti spezzo le gambe}" in italian.}. 
\section{GUT Baryogenesis at Preheating}

In models of slow-roll inflation, the universe is
dominated by the potential energy density of a scalar field known as
the {\it inflaton}.  Inflation ends when the kinetic energy density of
the inflaton becomes larger than its potential energy density.  At
this point the universe might be said to be frozen: any initial
entropy in the universe was inflated away, and the only energy was in
cold, coherent motions of the inflaton field.  Somehow this frozen
state must be transformed to a high-entropy hot universe by
transferring energy from the inflaton field to radiation.  

In
the simple chaotic inflation model  the potential is assumed
to be $V(\phi) = M_\phi^2\phi^2/2$, with $M_\phi\sim 10^{13}$GeV in
order to reproduce the observed temperature anisotropies in the
microwave background. In the old reheating (defrosting) scenario, the inflaton
field $\phi$ is assumed to oscillate coherently about the minimum of
the inflaton potential until the age of the universe is equal to the
lifetime of the inflaton.   Then the inflaton decays, and the decay
products thermalize to a temperature $T_F\simeq 10^{-1}
\sqrt{\Gamma_\phi M_{\rm P}}$, where $\Gamma_\phi$ is the inflaton
decay width, and $M_{\rm P}\sim 10^{19}$ GeV is the Planck mass.   

In supergravity-inspired scenarios, gravitinos have a mass of order a
TeV and a decay lifetime on the order of $10^5$s. If gravitinos are
overproduced after inflation and decay after the epoch of
nucleosynthesis, they would modify the successful predictions of
big-bang nucleosynthesis.  This can be avoided if the temperature
$T_F$ is smaller than about $10^{11}$ GeV (or even less, depending on
the gravitino mass). 

In addition to entropy, the baryon asymmetry must be created after
inflation.  There are serious obstacles facing any attempt to generate
a baryon asymmetry in an inflationary universe through the decay of
baryon number ($B$) violating bosons of Grand Unified Theories.  
The most  tedious problem is the low value of $T_F$ in the old scenario.
Since the unification scale is expected to be of order $10^{16}$GeV,
$B$ violating gauge and Higgs bosons (referred to generically as
``$X$'' bosons) probably have masses greater than $M_\phi$, and it
would be kinematically impossible to produce them directly in $\phi$
decay or by scatterings in a thermal environment  at temperature $T_F$.

However,  
reheating may differ significantly from the above simple picture~\cite{explosive}.  In
the first stage of reheating, which was called ``preheating''
effective dissipational dynamics and explosive particle production
even when single particle decay is kinematically forbidden. 
A crucial observation for baryogenesis is that even particles with
mass {\it larger} than the inflaton mass, $M_X\sim 10 \:M_\phi$, may be produced during preheating~\cite{explosive} by coherent effects provided that  a coupling to the inflaton field of the type $|X|^2 \phi^2$ is present.
A fully non-linear calculation of the   amplitude of perturbations $\langle X^2\rangle$ at the end of the broad resonance regime has been recently done~\cite{KT} revealing that  $\langle X^2\rangle$  may be as large as $10^{-10}
\:M_{\rm P}^2$. Since the  value of the inflaton field $\phi_i$ at the beginning of preheating is of order of $10^{-2}\:M_{\rm P}$, one may assume
that  the first step in reheating is to convert a fraction
$\delta\sim 10^{-4}$ of the inflaton energy density into a background of
baryon-number violating $X$ bosons.  They can be produced even if the
reheating temperature to be established at the subsequent stages of
reheating is much smaller than $M_X$. Here we see a significant
departure from the old scenario.  In the old picture production of $X$
bosons was kinematically forbidden if $M_\phi< M_X$, while in the new
scenario it is possible as the result of coherent effects. The
particles are produced out-of-equilibrium, thus satisfying one of the
basic requirements to produce the baryon asymmetry.
The next step in reheating is the decay of the $X$ bosons.  We assumed
that the $X$ decay products rapidly thermalize~\cite{paper}.  It is only after this
point that it is possible to speak of the temperature of the universe.
Moreover, we assumed that decay of an
$X$--${{\overline{X}}}$ pair produces a net baryon number $B/\epsilon$ (where $\epsilon$ is the CP-violating factor),
as well as entropy~\cite{paper}.   We have numerically integrated the Boltzmann equations describing the temporal evolution of the baryon number, of the energy density of $X$-particles  and  of the inflaton field. 
 Since the
number of $X$ bosons produced is proportional to $\delta$, the final
asymmetry is proportional to $\delta$ and we have  noted~\cite{paper} that $B/\epsilon \sim 10^{-9}$ can be
obtained for $\delta$ as small as $10^{-6}$.

Our scenario is based on several assumptions about the structure of
the theory, but the feeling is that  baryon number
generation may be  relatively efficient.
Within uncertainties of model parameters, the value of $\epsilon$,
etc., the present $B\sim10^{-10}$ may arise from GUT baryogenesis
after preheating.  Of course, additional work is needed to implement
the ideas discussed above in the context of a more realistic model and a complete numerical analysis able to decribe the dynamics from the end of inflation down to the final baryon asymmetry production through the preheating era is urgently called for~\cite{prep}.
\section*{Acknowledgments} I would like to thank my collaborators A. Linde and R. Kolb from whom I am still learning so much. I would also like  to thank Eleonora Riotto for entertaining discussions. This work is supported by the DOE and NASA
under Grant NAG5--2788.

\end{document}